 \documentclass[smallabstract,smallcaptions]{dccpaper}

\usepackage{epsfig}
\usepackage{citesort}
\usepackage{amsmath}
\usepackage{amssymb}
\usepackage{color}
\usepackage{url}
\usepackage{subfigure}
\usepackage[linesnumbered,ruled,noend,vlined, scleft,nofillcomment]{algorithm2e}

\newlength{\figurewidth}
\newlength{\smallfigurewidth}

\begin{document}

\title
{\large
\textbf{Two-Dimensional Block Trees\thanks{\scriptsize{Funded in part by European Union’s Horizon 2020  Marie Sk{\l}odowska-Curie grant agreement No 690941; MINECO  (PGE and FEDER) [TIN2016-78011-C4-1-R;TIN2013-46238-C4-3-R]; CDTI, MINECO [ITC-20161074;IDI-20141259;ITC-20151247]; Xunta de Galicia (co-founded with FEDER) [ED431G/01; ED431C 2017/58]; and Fondecyt Grants 1-171058 and 1-170048, Chile. }	}}
}

\author{%
Nieves R. Brisaboa$^{\ast}$, Travis Gagie$^{\dag}$,  Adri\'an G\'omez-Brand\'on$^{\ast}$, and Gonzalo Navarro$^{\ddag}$\\[0.5em]
{\small\begin{minipage}{\linewidth}\begin{center}
\begin{tabular}{ccc}
$^{\ast}$Database Laboratory & $^{\dag}$EIT & $^{\ddag}$Dept. of Computer Science \\
Universidade da Coru\~na & Diego Portales University & University of Chile \\
A Coru\~na, Spain & Santiago, Chile & Santiago, Chile\\
\url{brisaboa@udc.es} & \url{travis.gagie@mail.udp.cl} &\url{gnavarro@dcc.uchile.cl}\\
\url{adrian.gbrandon@udc.es} & &
\end{tabular}
\end{center}\end{minipage}}
}

\maketitle
\thispagestyle{empty}

\begin{abstract}
The Block Tree (BT) is a novel compact data structure designed to compress sequence collections. It obtains compression ratios close to Lempel-Ziv and supports efficient direct access to any substring. The BT divides the text recursively into fixed-size blocks and those appearing earlier are represented with pointers. On repetitive collections, a few blocks can represent all the others, and thus the BT reduces the size by orders of magnitude. In this paper we extend the BT to two dimensions, to exploit repetitiveness in collections of images, graphs, and maps. This two-dimensional Block Tree divides the image regularly into subimages and replaces some of them by pointers to other occurrences thereof.
We develop a specific variant aimed at compressing the adjacency matrices of Web graphs, obtaining space reductions of up to 50\% compared with the $k^2$-tree, which is the best alternative supporting direct and reverse navigation in the graph.

\end{abstract}

\Section{Introduction}

In many applications, image collections contain identical sub-images, for example two-dimensional slices of three-dimensional scans, video frames, and periodical sky surveys. This is an important source of redundancy that can be exploited for compression. Such an approach is two-dimensional Lempel-Ziv\cite{lempel1986compression} (2D-LZ), which stores only the first occurrence of each sub-image on the dictionary and the others are represented as pointers to the reference. However, 2D-LZ does not support efficient random access to individual images or arbitrary regions thereof. This is a relevant problem when storing large image collections in compressed form. Some proposals \cite{pajarola1996spatial,ageenko2000lossless} provide direct access by splitting the image into different partitions, and solving range queries by decompressing only those parts that intersect with the queried region. This induces, however, a tradeoff between extraction time and compression ratio, driven by the size of those partitions.

Other data like matrices, maps, and graphs, are also modeled as images and may contain similar areas. A particular case of repetitive two-dimensional data are {\em Web graphs}, which are directed graphs of pages pointing to other pages on the Web. The adjacency matrix of a Web graph can be seen as a bilevel image, where the link between pages $a$ and $b$ is represented with a 1 at position $(a,b)$, which has a 0 otherwise. Web graphs are sparse, so this matrix has large zones of 0s and a few 1s.

Since the adjacency matrix is huge and needs efficient random access, the design of compact data structures to represent Web graphs is a relevant topic. A well-known such structure is the $k^2$-tree \cite{ktree}. The $k^2$-tree is very efficient at representing large zones of 0s of the adjacency matrix and supporting direct and reverse neighbor queries. While there are other representations that exploit other properties of Web graphs (locality, similarity of adjacency lists, etc.) \cite{HNkais13,grabowski2011merging,boldi2004webgraph}, the $k^2$-tree offers the best space-time trade-off when considering both direct and reverse neighbor queries. However, the $k^2$-tree does not directly exploit repetitiveness.

The $k^2$-tree is, in essence, a compressed quad-tree of the matrix. Bille et al.~\cite{bille2015compressed} show how to replace subtrees of such structure by pointers to other identical subtrees, thereby turning the tree into a DAG. This, however, requires that two fixed subtrees are identical, which likely misses most repeated submatrices. We seek a way to capture such repetitiveness, while retaining fast queries.

A recent compact data structure called Block Tree (BT) \cite{belazzougui2015queries} compresses repetitive collections of (one-dimensional) strings. It obtains compression ratios close to Lempel-Ziv \cite{ziv1977universal} while supporting efficient direct access to any substring. The BT overcomes the inability of Lempel-Ziv in providing direct access by imposing a regular structure to the targets of string copies. The BT reduces the size of repetitive collections by orders of magnitude.

In this paper we extend the BT to two dimensions. The result, {\em Two-Dimensional Block Tree (2D-BT)}, can replace any subtree by a pointer to an {\em arbitrary} area where it is repeated, and still supports fast extraction of subimages. We then adapt the 2D-BT to Web graph adjacency matrices, by combining it with $k^2$-trees in order to exploit sparseness as well. Our experimental results on Web graphs show that the 2D-BT is typically 1.5--2 times smaller than the $k^2$-tree at the price of being 3--6 times slower. Although we develop the variant adapted to Web graphs in detail, the general 2D-BT can also be applied to images, matrices, maps, and other kinds of graphs.

\Section{Background}
\SubSection{$k^2$-tree}

The $k^2$-tree is a compact data structure designed to compress adjacency matrices of size $n^2$. It exploits the compression of large zones of 0s and the clustering of 1s.

The structure corresponds to a $k^2$-ary tree built by recursively splitting the adjacency matrix into $k^2$ submatrices of the same size. The construction algorithm first subdivides the whole adjacency matrix into $k^2$ submatrices of size $n^2/k^2$. These submatrices are sorted in row-major order, left to right and  top to bottom. The algorithm then checks each submatrix, adding a 1 as a child of the root if there is some 1 inside the submatrix, and otherwise adding a 0. The process continues recursively, splitting the submatrices with 1s into other $k^2$ submatrices. A submatrix is not split further when it is full of 0s or corresponds with a cell of the adjacency matrix. 

\begin{figure}[t]
\begin{center}
\epsfig{width=0.8\textwidth,file=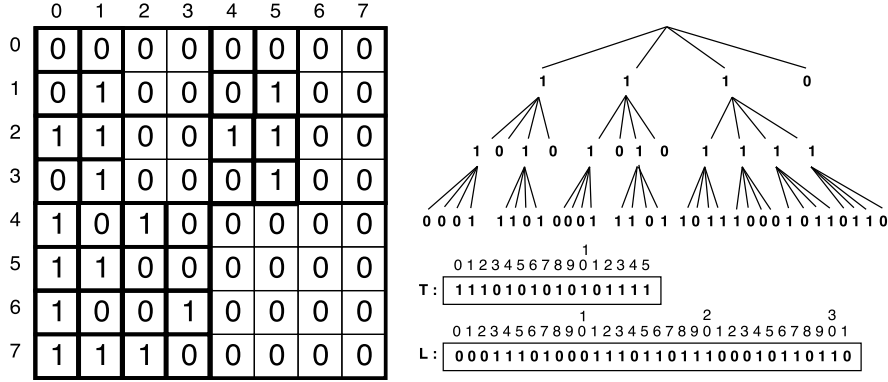}
\end{center}
\caption{\label{fig:k2tree}%
The $k^2$-tree of an adjacency matrix.}
\end{figure}

As shown in Figure \ref{fig:k2tree}, the $k^2$-tree is represented with two bitvectors: \textit{T} and \textit{L}. Bitvector $T[0..]$ stores all the bits of the $k^2$-tree except the last level. The bits are placed following a levelwise traversal of the tree. Bitvector $L[0..]$ stores the bits of the last level of the $k^2$-tree and each bit is the value of a cell.

It is possible to obtain any cell, row, column, or region of the matrix very efficiently, by traversing the appropriate subtrees. The traversal is simulated with \textit{rank} operations on the bitvector \textit{T}, where $rank_b(T,p)$ is the number of occurrences of bit $b \in [0, 1]$ up to position $p$. For example, given a value 1 at position $p$ in \textit{T}, its $k^ 2$ children will start at position
$p'=rank_1(T, p) \times k^2$ of $T$:$L$, so that their range is $children(p)=[p'..p'+k^2-1]$. Similarly, the parent of a position $p$ in $T$:$L$ is $parent(p)=select_1(T, \lfloor p/k^2 \rfloor)$, where $select_b(T,j)$ is the position of the $j$-th $b$ in \textit{T}. Both \textit{rank} and \textit{select} are implemented in constant time using $o(|T|)$ further bits \cite{clarktrees}.

\SubSection{Block Trees}

The Block Tree (BT) \cite{belazzougui2015queries} of a string $S[1..n]$ over an alphabet $[1..\sigma]$ uses $O(z \lg(n/z)\lg n)$ bits, where $z$ is the number of phrases produced by the Lempel-Ziv parse. We describe a simplified variant that is suitable for our purposes. We define a parameter $r$ as the arity of the BT. We start at level $l=0$, splitting $S$ into $r$ blocks of size $n/r$. Each of these blocks corresponds to a node of the tree. The nodes can be of two types: internal or leaves. A node $v$ is marked as internal if it contains or overlaps the first occurrence of the substring of a later block (or of itself). Otherwise, the node is a leaf. Each leaf stores a pointer $ptr$ to the node, or the first of the two nodes, that contain the leaf substring, and an offset \textit{off} that indicates the starting position of the occurrence within the pointed block. Once the first level is built, we split the internal nodes into $r$ of size $n/r^2$ and continue replacing each node's substring by a pointer wherever it first occurs in the same tree level. We recursively repeat these steps for each level until storing the substring is cheaper than storing the pointers and offsets.

To specify the type of the nodes, we use for each level $d$ a bitmap $D_d$ whose 0s represent the leaves of $d$ and the 1s are the internal nodes. A $rank$ structure is built over each bitmap $D_d$. Thus, running $j=rank_0(D_d,p)$ we can compute efficiently how many leaves are there up to $p$. The pointers of the leaves are stored into an array $P_d[0..]$ per depth $d$, so that $P_d[j-1]$ is the pointer of the $j$-th leaf. The offsets are similarly stored in arrays $O_d$. Instead, the 1s in $D_d$ are used to map the internal nodes to their position in level $d+1$, where only the internal nodes of level $d$ exist and are stored contiguously. 

The symbol at $S[i]$ can be found with a top-down traversal of the BT, going to the child node that contains $S[i]$, then the grandchild with $S[i]$, and so on. If the node is a leaf at the last level, we can access $S[i]$. Otherwise, the leaf has a pointer $i'$ to the internal node that contains the start of the first occurrence of the substring and an offset \textit{off}. Thus we translate $i$ to $i'= i +$\textit{off} and look for $S[i']$ instead. The BT guarantees that the node containing $i'$ is internal, so we can now descend. In total, any $S[i]$ is obtained in time $O(\log_r n)$

\Section{Two-Dimensional Block Trees}

Our new structure, Two-Dimensional Block Tree (2D-BT), can be seen as an extension of a Block Tree to two dimensions. It is designed to compress two-dimensional elements like matrices, images, or graphs. We first present a general 2D-BT structure. Then, we introduce a specific 2D-BT variant to compress Web graphs. This is a hybrid with the $k^2$-tree that exploits the clustering of 0s and, at the same time, the repetitiveness of the adjacency matrix.

\SubSection{Conceptual description}

Given a matrix $M$ of size $|M|=n^2$ over an alphabet $[1..\sigma]$, the matrix is subdivided into $k^2$ submatrices of size $n^2/k^2$. Each of these submatrices is called a block and represents a node of the 2D-BT. The nodes can be classified into internal or leaves. Consider any submatrix order, such as the row-major one used by the $k^2$-tree. Then nodes whose submatrix overlaps the first occurrence of a block (including themselves) are internal nodes; the others are leaves. The submatrix of any leaf node is said to be the {\em target} of a copy, whose {\em source} is its first occurrence. A  source may overlap up to four adjacent blocks. Each leaf stores a  pointer $ptr$ to the top-left block that includes its source and two offsets $O_x$ and $O_y$, one by axis, where the source starts in that block. Once the first level is built, we split the internal nodes into $k^2$ new nodes, and add them as children of the corresponding internal node. This step is repeated recursively until storing a pointer and its offsets is more expensive than storing the submatrix of a node. At this point, the submatrix content is stored verbatim. The 2D-BT has a maximum height of $height = \log_kn$.

To handle, in particular, Web graphs, we specialize this general 2D-BT structure so as to exploit clustering and sparseness, not only repetitiveness. We regard the adjacency matrix as a binary image. We define a new kind of leaf called {\em empty node}, which represents a block of all 0s. Therefore, leaves in this 2D-BT may be empty nodes, pointers to sources, or last-level nodes storing individual cells.

\begin{figure}[t]
\begin{center}
\epsfig{width=0.8\textwidth,file=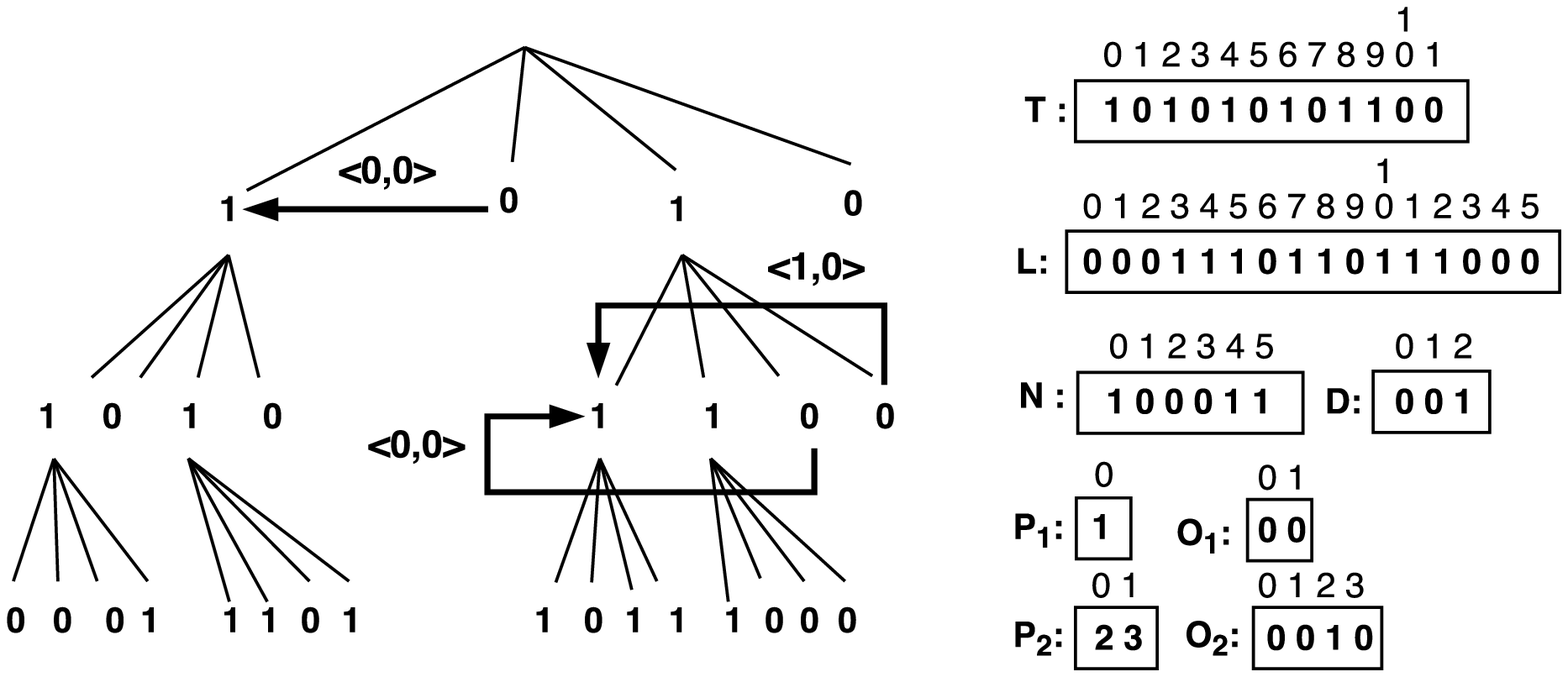}
\end{center}
\caption{\label{fig:2dbt}%
The 2D-BT, for $k=2$, of the adjacency matrix of Figure \ref{fig:k2tree}}
\end{figure}

An example of a 2D-BT for Web graphs, with $k=2$, can be observed on the left of Figure \ref{fig:2dbt}. It is built over the adjacency matrix of Figure \ref{fig:k2tree}.
The nodes with a 1 specify an internal node and they are divided into $k^2$ nodes in the next level. Empty nodes are represented with a 0, and leaf nodes with a 0 and a pointer. The pointers are arrows to an earlier node in the same level and they are labeled with an offset $\langle O_x, O_y \rangle$. Only internal nodes are further split. An example of a node with pointer is the last one of the second level, which points to the fifth node with $O_x=1$ and $O_y=0$. Since the target submatrix is $[(2,6), (3,7)]$\footnote{The submatrix is represented with the top-left and the bottom-right corners in $(x, y)$ format.} and the source is $[(1,4), (2,5)]$, we add the offsets $\langle 1, 0 \rangle$ to the coordinates of the pointed node, $[(0,4), (1,5)]$.

The existence of empty nodes makes less obvious whether it is convenient or not to store a pointer to replace a target. We must consider its representation using empty nodes (i.e., as a $k^2$-tree) and compare it with the space used by the pointer and offsets. An area that is better represented as a $k^2$-tree should not be used as a source, so its first occurrence should not be used to mark nodes as internal.

\SubSection{Data structure}

As shown on the right of Figure \ref{fig:2dbt}, we represent and navigate the 2D-BT much as a $k^2$-tree, using bit arrays {\em T} and {\em L}. The difference is that some leaves in {\em T} (0s) represent empty nodes and others represent pointers. These are matched with 0s and 1s, respectively, in another bit array $N[0..m-1]$, where {\em T} has $m$ 0s. Then a leaf $T[p]=0$ is a pointer iff $N[rank_0(T,p)-1]=1$.

The pointers and offsets are stored in different arrays, one per depth. For each depth $d$, except the last one where there are no pointers, an array $P_d$ stores the pointers $ptr$ of that depth. The values $P_d$ are stored as backward offsets: the position in $T$ pointed from $T[i]$ by $P_d[q]$ is $i-P_d[q]$. We store only the maximum of the bits needed for the cells of each array $P_d$. The offsets are stored into arrays $O_d$. The even positions are the $x$-offsets $O_x$, and the odd positions are $y$-offsets $O_y$. For the cells in $O_d$ we only need $\lceil \log_k(n/k^d)\rceil=\log_k(n)-d$ bits. Let a leaf $p$ at level $d$ be a pointer, i.e., $T[p]=0$ and $N[p'-1]=1$ with $p'=rank_0(T,p)$. The pointer is stored at position $P_d[rank_1(N,p')-D[d]]$, where $D$ is a small array accumulating the number of positions in previous levels. Its offsets are accessed similarly.

In Figure \ref{fig:2dbt}, at position $p=11$ ($p'=rank_0(T,11)=6$) and $d=2$, there is a pointer because $T[11]=0$ and $N[5]=1$. We know that the number of cells in arrays $P_d$ in previous levels is $D[2]=1$. We then compute $q=rank_1(N,5)-D[2]=3-1=2$, thus the pointer is at $P_d[2-1]$. Therefore, $ptr=11-P_2[1]=8$ and offsets $\langle 1,0 \rangle$ because $O_2[2]=1$ and $O_2[3]=0$.

\SubSection{Access to a region}

\begin{figure}[t]
  \begin{minipage}[t]{.565\textwidth}
    \begin{algorithm}[H]
		\caption{{\bf access}({$R$, $p$, $d$, $r$, $c$})\label{alg:access}}
            \If{$d = height$}{
            	\If(){$L[p - |T|] = 1$}{ 
                	$result[r][c]=1$
                }
            }
            \Else{
            	\If(){$p\ is\ internal$}{ 
                    \For(){$p'$ in $children(p)$}{
                        $Q \gets region(p')$\\
                        \If(){$Q \cap R \neq \emptyset$}{
                            $\textbf{access}(Q \cap R,p', d+1,c,r)$
                        }
                        $r,c \gets next()$\\
                    }
                }
                \Else{
                	\If(){$p\ is\ a\ pointer$}{
                	$p_{ptr}, O_x, O_y \gets pointer(p)$\\
                    $R', p', d' \gets \textbf{back}(p_{ptr}, O_x, O_y,d)$\\
                    $\textbf{access}(R', p', d', r, c)$
                }
                }
                
             }
	\end{algorithm}
  \end{minipage}%
  \hfill\vrule\hfill
  \begin{minipage}[t]{.425\textwidth}
    \begin{algorithm}[H]
      \caption{{\bf back}({$p_{ptr}$,$O_x$,$O_y$,$d$})\label{alg:back}}
      		$R' \gets region(p_{ptr})+(O_x, O_y)$\\
            \While{$R' \nsubseteq region(p_{ptr})$}{
            	$p_{ptr} \gets parent(p_{ptr})$\\
                $d \gets d-1$
            }
            \Return $R'$, $p_{ptr}$, $d$
            \vspace{5.1cm}
    \end{algorithm}
  \end{minipage}
\end{figure}
Given a region $R = [(x_{min}, y_{min}), (x_{max}, y_{max})]$ the operation $access(R, 0, 0, 0, 0)$ in Algorithm \ref{alg:access} writes the values of the region into an initially zeroed matrix $result$. These values are retrieved by  descending the 2D-BT from the root through all the nodes $p$ whose area, $region(p)$, intersect $R$, until finding the 1s at the tree leaves. The algorithm uses two non-obvious notations:
\begin{itemize}
\item $region(p)$ is the region $Q$ covered by the node at position $p$ in $T$. The region is easily maintained as we traverse the tree. At depth $d$, given the top-left corner $\langle x_p, y_p\rangle$ and size $side \times side$ of $Q$, the $i$-th child of $p$ ($i=0...k^2-1$) starts at $x$-offset $x_p+ (i \!\!\mod k) \cdot (side/k)$ and $y$-offset $y_p+(\lfloor i / k \rfloor) \cdot (side/k)$. Similarly, if $p$ is the $i$-th child of $p'=parent(p)$, then the region $p'$ starts at $x$-offset $x_p - (i \mod k) \cdot side$ and $y$-offset $y_p - (\lfloor i / k \rfloor) \cdot side$.

The only case where we cannot obtain $region(p)$ from the navigation is for $region(p_{ptr})$ in the first line of Algorithm \ref{alg:back}. In fact, instead of explicitly computing $region(p)$, we always use $R$ relative to the current $region(p)$. Thus, when we move from region $R$ relative to $p$ to region $R'$ relative to $p_{ptr}$ in the first line of Algorithm \ref{alg:back}, all we have to do is $R' \gets R + (O_x,O_y)$.

\item $r, c \gets next()$: obtains the top-left position of the next child in the $result$ matrix, in row-major order. It is computed analogously to the child region. 
\end{itemize}

The most interesting part of the algorithm is how we handle the pointers. Given the target node $p_{ptr}$ and the offset $\langle O_x, O_y \rangle$, the problem is how to find the other (up to) 3 nodes that contain parts of the target. Algorithm \ref{alg:back} goes up until finding an ancestor at depth $d'$ that contains the target. Then the process continues from that ancestor. Note that, although we are going higher in the 2D-BT, such recursive call will eventually go back to level $d$ only for the (up to) 4 desired nodes, entering into (up to) 4 children at level $d'$ and then into only one child in levels $d'+1$ to $d-1$.

Therefore, the cost to find the desired neighbors of $p_{ptr}$ is $O(d-d')$. On average, the probability of the target not being contained in one node at level $d-h$ is $< 2/k^h$; therefore on average $d-d'$ is $O(1)$. 
Still, we can obtain an $O(h)$ time guarantee by storing two pointers to the node at the right and the bottom of all nodes at depth $d\leq height-h$. This requires a total extra space of $O(\frac{n^2}{k^{2h}}\log n)$ bits.

\SubSection{Construction}

In order to build a 2D-BT, we need to efficiently identify the first occurrence of the submatrix of every block. We use an algorithm based on the technique of Karp and Rabin \cite{karp1987efficient}, which obtains a fingerprint of a string. Given a substring $S[i,j]$, the Karp-Rabin algorithm computes the fingerprint, called the \textit{KR-fingerprint}, in time $O(j-i+1)$. However, given the fingerprint of $S[i,j]$, the fingerprint of $S[i+1, j+1]$ is computed in time $O(1)$. We extend this scheme to two dimensions, in the style of Bird \cite{bird1977two} and Baker \cite{baker1978technique}.

Let $side$ be size of a block side in a given level. In the first phase we compute a matrix $K\!Rmatrix$, where $K\!Rmatrix[row][col]$ is the \textit{KR-fingerprint} of the substring $M[row,...,row+side-1][col]$. In the second phase we compute a second $K\!Rmatrix'$, where $K\!Rmatrix'[row][col]$ is the KR-fingerprint of $K\!Rmatrix[row][col,...,col+side-1]$, thus it is a KR-fingerprint of the submatrix $M[row,...,row+side-1][col, ...,col+side-1]$. This algorithm takes linear time on the size of $M$, $O(n^2)$, and $O(n^2)$ space.

After computing $K\!Rmatrix'$, we add the fingerprints $K\!Rmatrix'[i\times side][j\times side]$ to a hash table, for every $0 \leq i, j < n/side$. These are the fingerprints of all the possible targets in the current level. Associated with each KR-fingerprint in the hash table we store a list with the submatrices we found with that fingerprint, and for each distinct submatrix, a list of the targets containing that submatrix. Note that we must not insert targets whose $k^2$-tree representation is smaller than our pointers, since replacing them is more expensive than continuing the $k^2$-tree deployment.

Once the hash table is built, we go through the cells of $K\!Rmatrix'$ checking if each fingerprint is contained in the hash table. For every cell $h=K\!Rmatrix'[i][j]$ we find in the hash table, we check if $R=M[i,...,i+side-1][ j,...,j+side-1]$ is one of the submatrix contents associated with $h$. If it is, then $R$ is the first occurrence of all the associated targets. All those are removed from the hash table, and their nodes are converted into a leaf pointing to $R$. The (up to) 4 nodes covered by $R$ are marked to ensure they are internal nodes (in particular, if the target overlaps $R$, it will not be converted into a leaf). Further, we must exclude as possible source any area $R$ that overlaps nodes that have been converted into leaves.

\Section{Experimental Evaluation}

\begin{figure}[t]
\centering     
\includegraphics[width=0.7\textwidth]{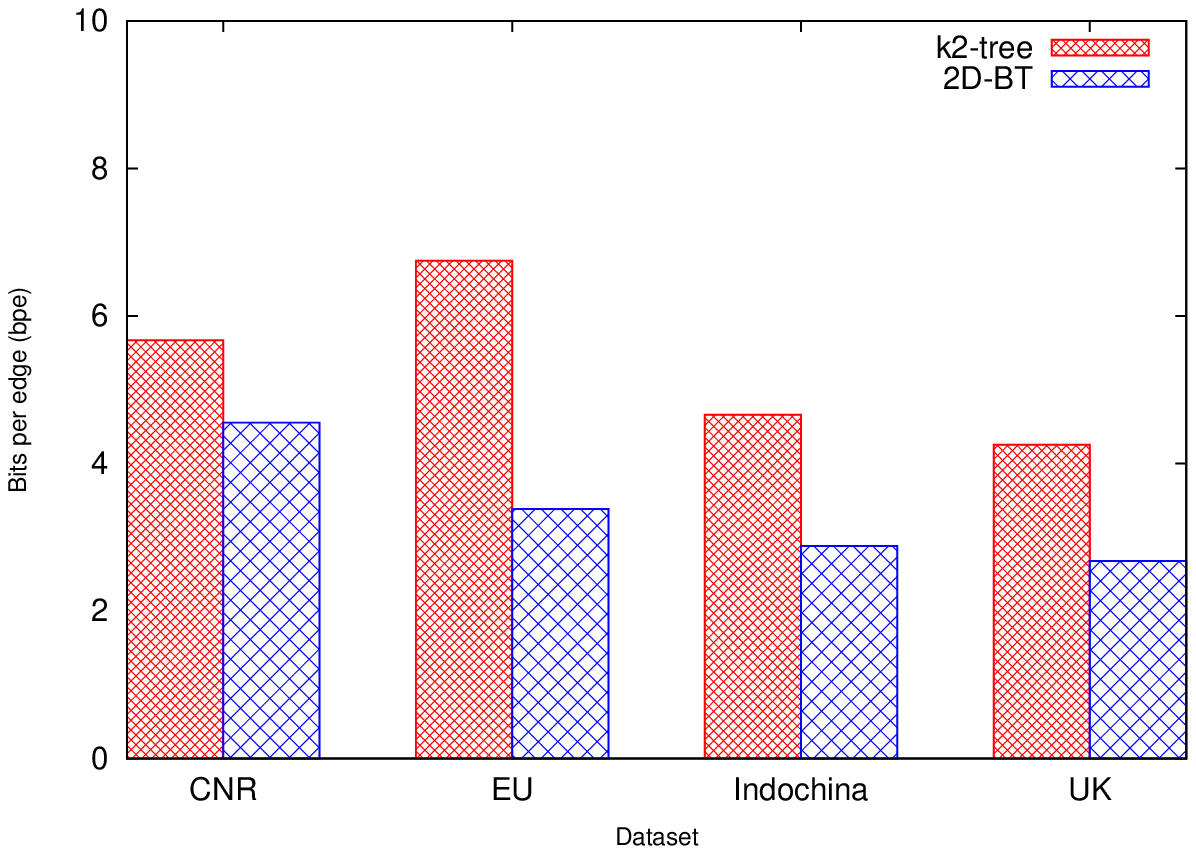}
\caption{The bits per edge (bpe) of $k^2$-tree and 2D-BT on four Web subgraphs.}
\label{fig:size}
\end{figure}
Two-Dimensional Block Trees were coded in C++,  using several data strucures from the SDSL library \cite{gbmp2014sea}. As a baseline, we include the SDSL implementation of $k^2$-trees. We used four real datasets of the \textit{WebGraph framework} \cite{boldi2004webgraph}: CNR (2000), EU (2005), Indochina (2002) and UK (2002). These are avaliable from the site \url{http://webgraph.dsi.unimi.it}. The experiments ran on an Intel\textsuperscript{\textregistered} Xeon\textsuperscript{\textregistered} CPU E5-2407 v2 @ 2.40GHz (8 cores) with 10MB of cache and 256GB of RAM, over SMP Debian 3.16.43-2 with kernel 3.16.0-4-amd64 (64 bits), using gcc 4.6.4 with \texttt{-O9}.

We use $k=2$ and compare the size of both structures and the average time to retrieve \textit{direct neighbors} and \textit{reverse neighbors}. Extracting the direct neighbors of a node $i$ is equivalent to retrieving the $i$-th row of the adjacency matrix, whereas its reverse neighbors correspond to the $i$-th column. Since our construction takes much time and memory space, we limited the number of nodes of our structure to  $100,000$ and collected the induced subgraphs.


\begin{figure}[t]
\centering     
\subfigure[\texttt{Dataset CNR}]
{\label{fig:CNR}\includegraphics[width=0.49\textwidth]{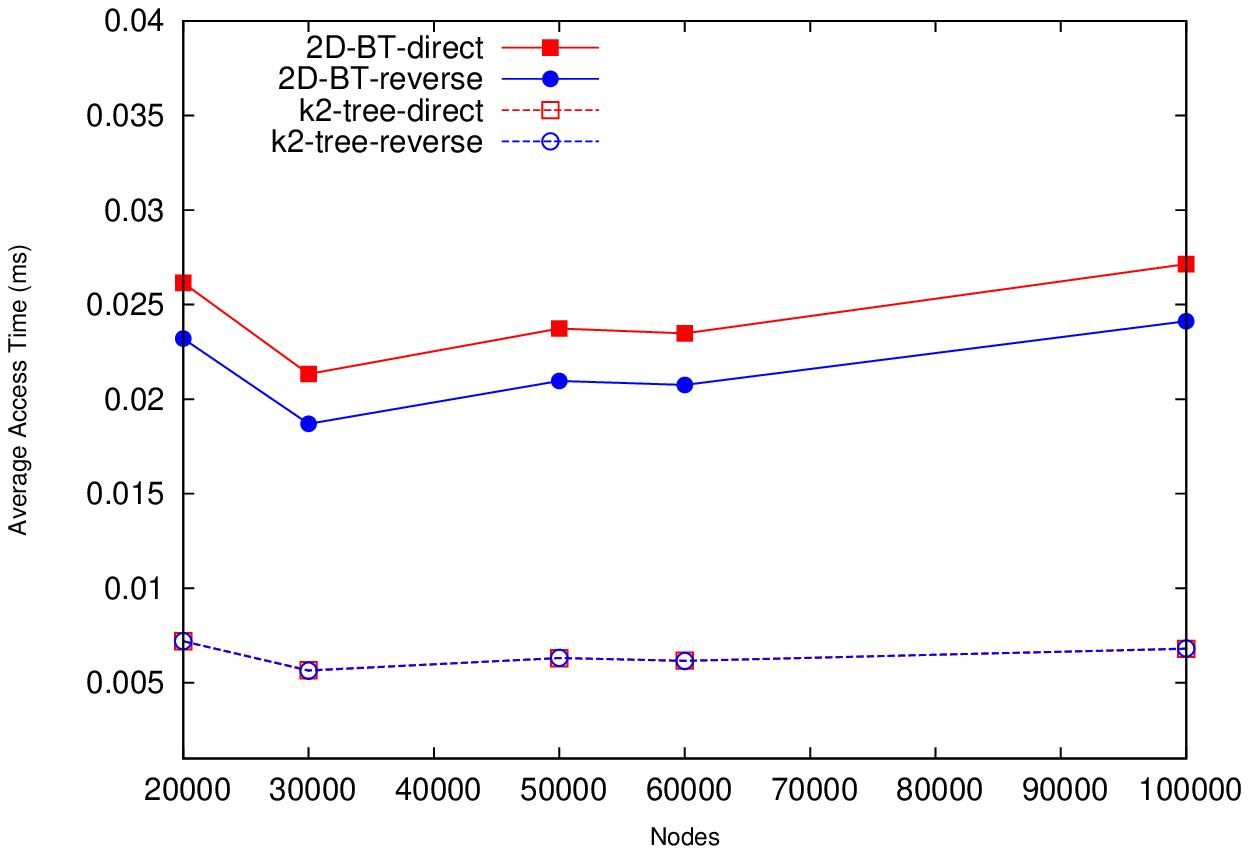}}
\subfigure[\texttt{Dataset EU}]
{\label{fig:EU}\includegraphics[width=0.49\textwidth]{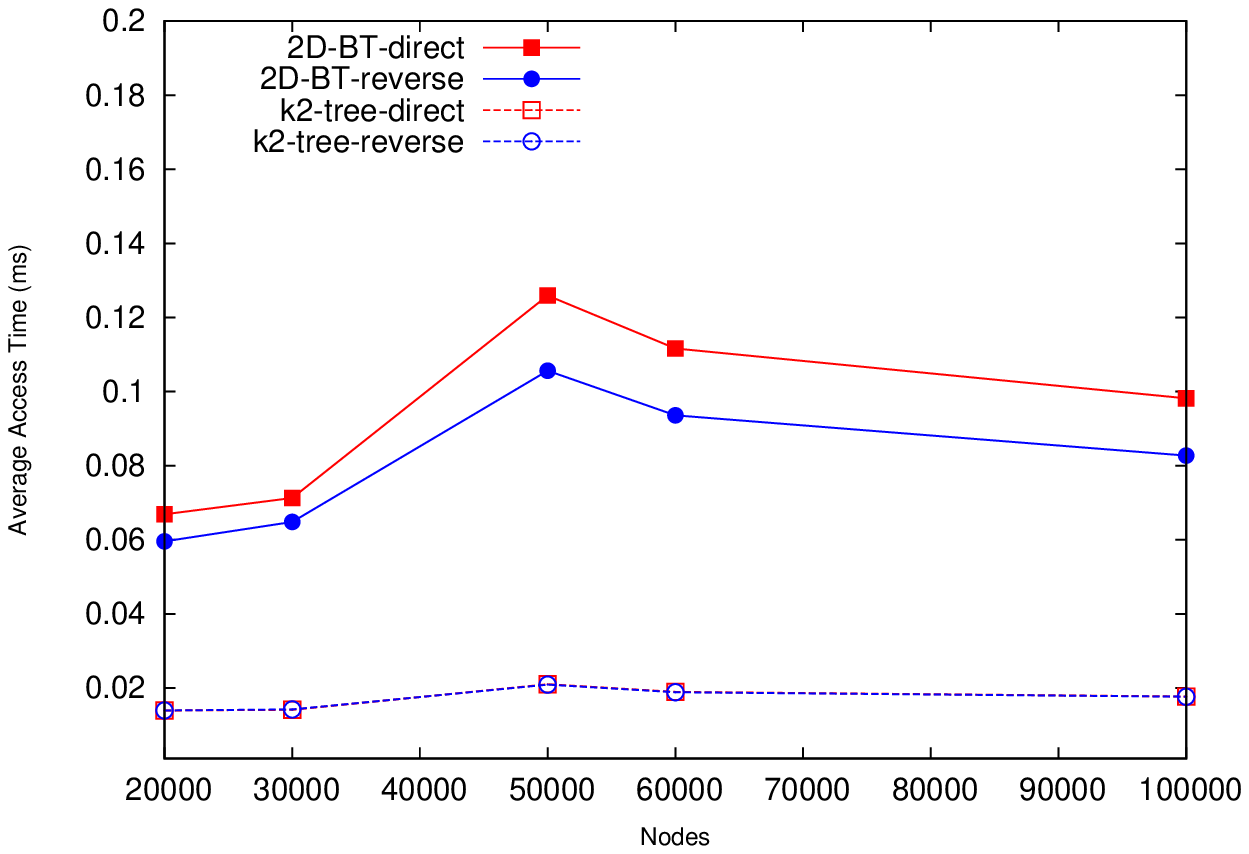}}
\subfigure[\texttt{Dataset Indochina}]
{\label{fig:Indochina}\includegraphics[width=0.49\textwidth]{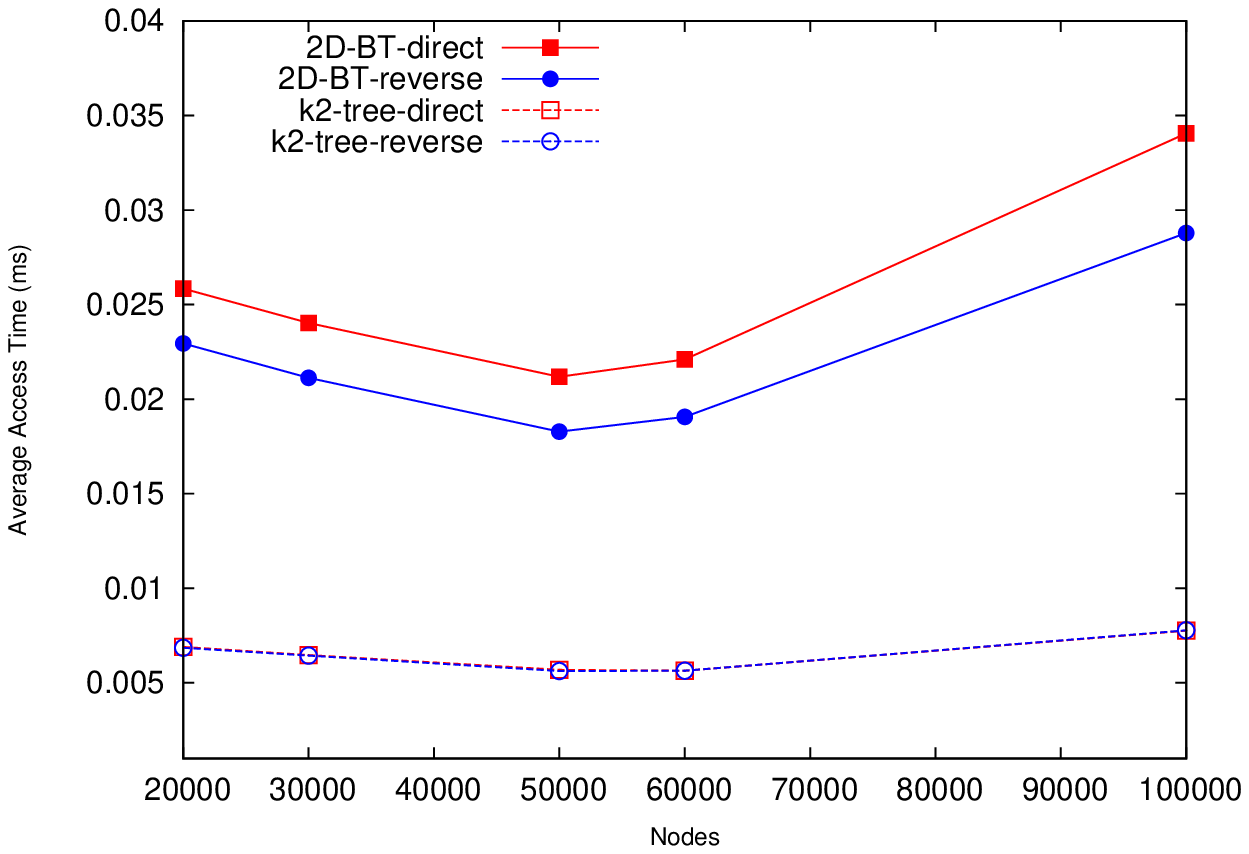}}
\subfigure[\texttt{Dataset UK}]
{\label{fig:UK}\includegraphics[width=0.49\textwidth]{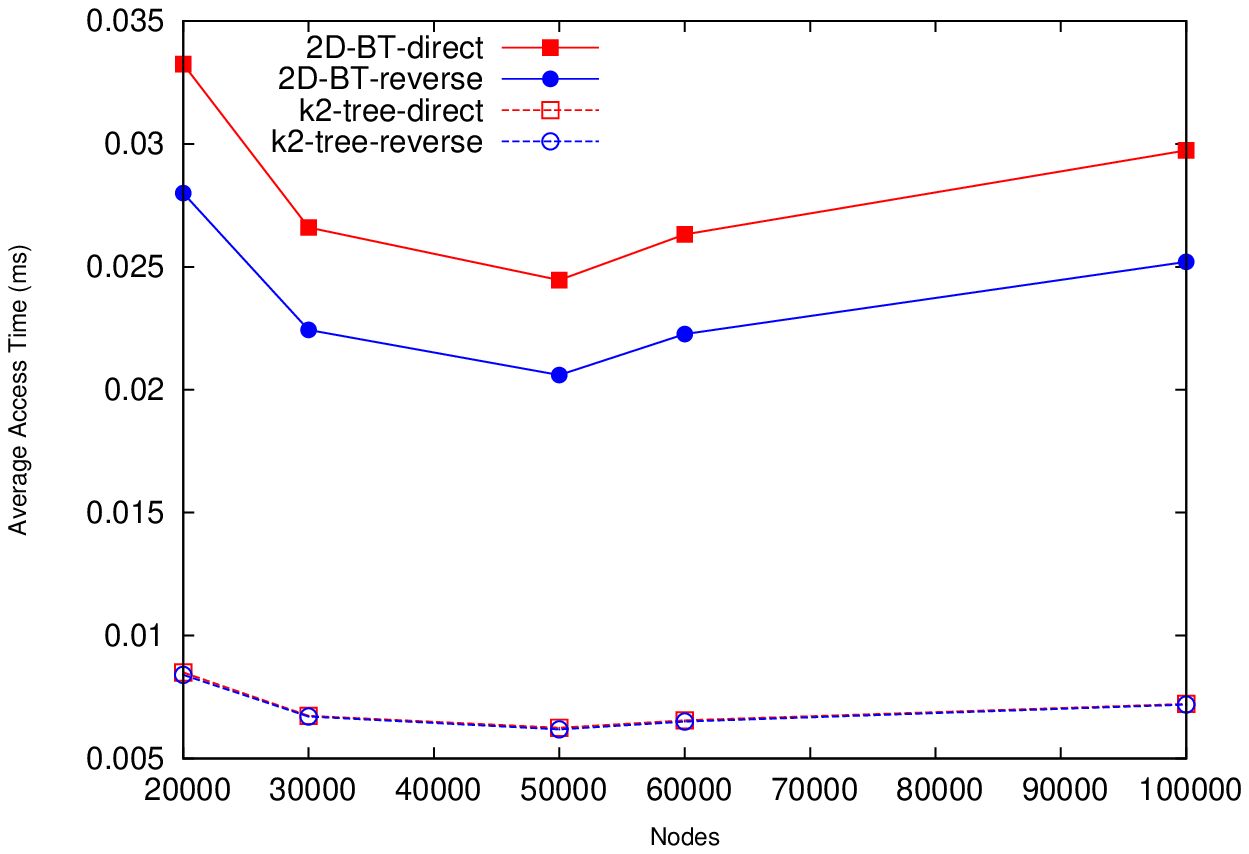}}
\caption{Average access times on the the four Web subgraphs. The times on $k^2$-trees are very similar and thus look superimposed.}
\label{fig:times}
\end{figure}

As we can observe in Figure \ref{fig:size}, the 2D-BT outperforms $k^2$-tree in space, since the 2D-BT exploits the repetitiveness of patterns, not only the clustering of zeros: the 2D-BT uses 50\%-65\% of the space in all datasets except CNR (80\%). 

Figure \ref{fig:times} shows the average access times to direct and reverse neighbors. We observe that the behavior is similar between both structures. Increasing the number of nodes of each dataset, the heights of the 2D-BT and the $k^2$-tree increase, and the average times are higher. Since traversing the 2D-BT is more expensive, our structure  obtains worst access times compared to $k^2$-tree. We observe that 2D-BTs are 3--6 times slower than $k^2$-trees. 


We measured the average difference $d-d'$ when using Algorithm \ref{alg:back}, to find out if the effect of finding the right ancestor is significative. Our experiments show that, on average, we go up only 1.5 levels, so this is not a concern in practice.

\Section{Conclusions and Future Work}

We have proposed a new structure that extends Block Trees \cite{belazzougui2015queries} to two dimensions, and combined them with $k^2$-trees \cite{ktree} to handle in particular Web graphs. On those, we obtained up to 50\% of the space of $k^2$-trees (the best structure that allows navigating the graph in both directions). The price is that we are 3--6 times slower. This price can be irrelevant when the lower space allows fitting the whole graph in a faster memory (e.g., RAM vs disk).

Our most immediate future work is to improve the construction, in order to handle full Web graph adjacency matrices. The current construction takes too much space and time. We plan to replace the 2D signature based scheme by a randomized method based on sampling positions in the submatrix. We also plan to explore other application areas where the values of the matrix display a good deal of repetitiveness.

\Section{References}
\bibliographystyle{IEEEtran}
\bibliography{refs}

\end{document}